\documentstyle[aps]{revtex}
\begin{document}
\draft
\author{{\bf H. Hern\'{a}ndez\thanks{
e-mail: hector@ciens.ula.ve}, }}
\address{{\it Laboratorio de F\'{\i}sica Te\'{o}rica, }\\
{\it Departamento de F\'{\i}sica, Facultad de Ciencias, }\\
{\it Universidad de Los Andes, M\'{e}rida 5101, Venezuela}\\
and\\
{\it Centro Nacional de C\'{a}lculo Cient\'{\i}fico}\\
{\it Universidad de Los Andes {\tt (CeCalCULA)},}\\
{\it Corporaci\'{o}n Parque Tecnol\'{o}gico de M\'{e}rida, }\\
{\it M\'{e}rida 5101, Venezuela}}
\author{{\bf L. A. N\'{u}\~{n}ez\thanks{
e-mail: nunez@ciens.ula.ve}}}
\address{{\it Centro de Astrof\'{\i}sica Te\'{o}rica,}\\
{\it Departamento de F\'{\i}sica, Facultad de Ciencias, }\\
{\it Universidad de Los Andes, M\'{e}rida 5101, Venezuela}\\
and\\
{\it Centro Nacional de C\'{a}lculo Cient\'{\i}fico}\\
{\it Universidad de Los Andes {\tt (CeCalCULA)},}\\
{\it Corporaci\'{o}n Parque Tecnol\'{o}gico de M\'{e}rida,}\\
{\it M\'{e}rida 5101, Venezuela}}
\author{and {\bf U. Percoco\thanks{%
e-mail: upercoco@ciens.ula.ve}}}
\address{{\it Centro de Astrof\'{\i}sica Te\'{o}rica, }\\
{\it Departamento de F\'{\i}sica, Facultad de Ciencias, }\\
{\it Universidad de Los Andes, M\'{e}rida 5101, Venezuela}}
\title{Nonlocal Equation of State \\
in General Relativistic\\
Radiating Spheres}
\date{May 1998}
\maketitle

\begin{abstract}
We show that under particular circumstances a general relativistic
spherically symmetric bounded distribution of matter could satisfy a
nonlocal equation of state. This equation relates, at a given point, the
components of the corresponding energy momentum tensor not only as function
at that point, but as a functional throughout the enclosed configuration. We
have found that these types of dynamic bounded matter configurations, with
constant gravitational potentials at the surface, admit a Conformal Killing
Vector and fulfill the energy conditions for anisotropic imperfect fluids.
We also present several analytical and numerical models satisfying these
equations of state which collapse as reasonable radiating anisotropic
spheres in general relativity.
\end{abstract}

\section{Introduction}

Our comprehension of the behavior of highly compact stars is intimately
related to the understanding of the physics at supranuclear densities.
Today, the properties of matter at densities higher than nuclear ( $\approx
10^{14}$ $gr/cm.^3$) are essentially unknown, although they must be oriented
by the experimental insight and extrapolations emerging from the ultra high
energy accelerators and firmly anchored to what is known from nuclear
physics \cite{Demianski85,ShapiroTeukolsky83}. Having this uncertainty in
mind, we shall explore what is allowed by the laws of physics, for a
particular equation of state, within the framework of the theory of General
Relativity and considering spherical symmetry.

We shall consider a spherically symmetric space-time that can be described
by the metric 
\begin{equation}
ds^2=h\,e^{4\beta }dT^2-\frac 1hdR^2-R^2d\Omega ^2\,,  \label{metrica}
\end{equation}
where the solid angle is $d\Omega ^2\equiv d\theta ^2+\sin ^2\theta d\phi
^2\,$ and both $\beta $ and $h$ depend on $T$ and $R$.

We have found that if in the above space-time (\ref{metrica}) an additional
restriction, 
\begin{equation}
h(T,R)\equiv 1-\frac{2m(T,R)}R={\rm C}(T)\,e^{-2\beta (T,R)},\,\quad {\rm %
with}\quad 0<{\rm C}(T)<1\ ,  \label{condicion}
\end{equation}
is considered, we can obtain the following relation between two of the
components of the corresponding Einstein tensor. i.e. 
\begin{equation}
{\bf G}_T^T+3{\bf G}_R^R+R\frac \partial {\partial R}\left[ {\bf G}_T^T+{\bf %
G}_R^R\right] =0\,.  \label{ecua5}
\end{equation}
Equation (\ref{ecua5}) can be integrated yielding 
\begin{equation}
{\bf G}_T^T=-{\bf G}_R^R-\frac 2R\,{\int }_0^R{\bf G}_R^R\ {\rm d}{\bar{r}\
+\ }\frac{{\cal K}(T)}R\ ,  \label{globalG}
\end{equation}
which, using the Einstein equations, can be re-written as 
\begin{equation}
{\bf T}_T^T=-{\bf T}_R^R-\frac 2R\,{\int }_0^R{\bf T}_R^R\ {\rm d}{\bar{r}\
+\ }\frac{{\cal K}(T)}R\ ,  \label{global}
\end{equation}
where ${\cal K}(T)$ is an arbitrary integration function. Obviously, from (%
\ref{ecua5}), we can also solve ${\bf G}_R^R$ in order to obtain an
expression equivalent to (\ref{global}), i.e. 
\begin{equation}
{\bf T}_R^R=-{\bf T}_T^T+\frac 2{R^3}\int_0^R\bar{r}^2{\bf T}_T^T\ {\rm d}%
\bar{r}\ -\frac{{\cal G}(T)}{R^3}\ .  \label{global2}
\end{equation}
Again, ${\cal G}(T)$ is an arbitrary integration function. Clearly, for
bounded matter distributions both functions ${\cal K}(T)$ and ${\cal G}(T),$
should emerge from the junction conditions to match the corresponding
exterior metric.

It can be easily appreciated that the above equations (\ref{global}) and (%
\ref{global2}) involve a non local relation between the components of the
energy momentum tensor. At a given instant of time, one of the component
(say ${\bf T}_T^T$ in (\ref{global})) is not only a function of the other ( $%
{\bf T}_R^R$ ) at that point but also its functional throughout the rest of
the configuration. In the case of radiating fluids, equations (\ref{global})
and (\ref{global2}) relate non-locally the hydrodynamic and radiation
physical variables.

Equation (\ref{global2}) could be further re-written as 
\begin{equation}
{\bf T}_R^R=-{\bf T}_T^T+\frac 23\left\langle {\bf T}_T^T\right\rangle -%
\frac{{\cal G}(T)}{R^3}\ ,  \label{averagemunu}
\end{equation}
with 
\begin{equation}
\left\langle {\bf T}_T^T\right\rangle =\frac 1{\frac{4\pi }3R^3}\int_0^R4\pi 
\bar{r}^2{\bf T}_T^T\ {\rm d}\bar{r}\ ,  \label{averagetensor}
\end{equation}
clearly the nonlocal term represents an average of the function ${\bf T}_T^T$
over the volume enclosed by the radius $R$. Moreover, equation (\ref
{averagemunu}) can be easily rearranged as 
\begin{equation}
{\bf T}_R^R=-\frac 13{\bf T}_T^T-\frac 23\ \left( {\bf T}_T^T-\left\langle 
{\bf T}_T^T\right\rangle \right) \ -\frac{{\cal G}(T)}{R^3}=-\frac 13{\bf T}%
_T^T-\frac 23\ {\bf \sigma }_{{\bf T}_T^T}-\frac{{\cal G}(T)}{R^3}\ ,
\label{sigmatmunu}
\end{equation}
where we have used the concept of statistical standard deviation ${\bf %
\sigma }_{{\bf T}_T^T}$ from the local value of the function ${\bf T}_T^T$ .
Therefore, if at a particular point within the distribution the value of the
function ${\bf T}_T^T$ gets very close to its average $\left\langle {\bf T}%
_T^T\right\rangle $ the relation between ${\bf T}_R^R$ and ${\bf T}_T^T$
becomes 
\begin{equation}
{\bf T}_R^R\approx -\frac 13{\bf T}_T^T\ -\frac{{\cal G}(T)}{R^3}\quad {\rm %
with\quad }{\bf \sigma }_{{\bf T}_T^T}\approx 0\ .  \label{sigma0}
\end{equation}

Physical insight can be gained by considering equations (\ref{global}) and (%
\ref{global2}) in their static limit and in terms of the hydrodynamic
density, $\rho (R)$ and pressure, $P_r(R)$. In this case, a static
spherically symmetric distribution leads to 
\begin{eqnarray}
\rho (R) &=&P_r(R)+\frac 2R\,{\int }_0^RP_r\,({\bar{r}})\ {\rm d}{\bar{r}\
+\ }\frac{{\cal H}}R\ ,\qquad {\rm or}\qquad  \label{globalestat} \\
P_r(R) &=&\rho (R)-\frac 2{R^3}\int_0^R\bar{r}^2\rho (\bar{r})\ {\rm d}\bar{r%
}\ +\frac{{\cal B}}{R^3}\ ;  \label{globalestat2}
\end{eqnarray}
with ${\cal H}\ $ and ${\cal B}$ are arbitrary integration constants. It is
clear that in equations (\ref{globalestat}) and (\ref{globalestat2}) a
collective behavior on the physical variables $\rho (R)$ and $P_r(R)$ is
also present, and could be interpreted as a nonlocal interrelation between
the energy density, $\rho (R),$ and the hydrodynamic pressure, $P_r(R),$%
within the fluid. Any change in the pressure (density) takes into account
the effects of the variations of the energy density (pressure) within an
entire volume.

Following the above line of reasoning we shall consider the static limit of
the equation (\ref{sigmatmunu}): 
\begin{equation}
P_r(R)={\cal P}(R)+2{\bf \sigma }_{{\cal P}(R)}+\frac{{\cal B}}{R^3}\ ,
\label{densaver}
\end{equation}
where we have set 
\begin{equation}
{\cal P}(R)=\frac 13\rho (R)\qquad {\rm and}\quad {\bf \sigma }_{{\cal P}%
(R)}=\left( \frac 13\rho (R)-\frac 13\left\langle \rho \right\rangle
_R\right) =\left( {\cal P}(R)-{\bar{{\cal P}}}(R)\right) ,
\label{sigmastatic}
\end{equation}
with 
\begin{equation}
\left\langle \rho \right\rangle _R=\frac 1{\frac{4\pi }3R^3}\int_0^R4\pi 
\bar{r}^2\rho (\bar{r})\ {\rm d}\bar{r}\ =\frac{M(R)}{\frac{4\pi }3R^3}.
\label{averagerho}
\end{equation}
In this limit the nonlocal integral term in (\ref{globalestat2}) represents
some kind of ``average density'' over the enclosed volume and, as far as the
value of this nonlocal contribution gets closer to the value of the local
density, the equation of state of the material becomes similar to the
typical radiation dominated environment, i.e. $P_r(R)\approx {\cal P}%
(R)\equiv \frac 13\rho (R)$.

It is in the above sense, that we are going to refer to (\ref{global}) or to
(\ref{global2}), as {\it Nonlocal Equation of State} ({\it NLES} from now
on) between the components ${\bf T}_T^T$ and ${\bf T}_R^R$ of the
corresponding energy momentum tensor.

Nowadays, the relevance of nonlocal outcomes on the mechanical properties of
fluids and materials are well known. There are many situations of common
occurrence wherein nonlocal effects dominate the macroscopic behavior of
matter in Modern Classical Continuum Mechanics and Fluid Dynamics. Examples
coming from a wide variety of classical areas such as: damage and cracking
analysis of materials, surface phenomena between two liquids or two phases,
mechanics of liquid crystals, blood flow, dynamics of colloidal suspensions
seem to demand more sophisticated continuum theories which can take into
account different nonlocal effects. This is a very active area in recent
material and fluid science and engineering (see \cite{Narasimhan93} and
references therein). Moreover, in radiating fluids when the medium is very
transparent and photons escape efficiently, occupation numbers of atomic
levels are, in general, no longer predicted by equilibrium statistical
mechanics for local values of the temperature and density. In this case, the
local state of a fluids is coupled by photon exchange to the state of the
material within an entire interaction volume. These types of fluids are
considered to be in {\it non-Local Thermodynamical Equilibrium} or non-LTE 
\cite{MihalasMihalas84}.

In this work only collapsing radiating configuration in General Relativity
will be studied. Static models consistent with {\it NLES} require a detail
analysis and will be considered elsewhere \cite{HernandezNunez98}. In what
follows, we shall explore four major questions concerning the {\it NLES} in
collapsing spherical sources in General Relativity, which can be stated as
follows:

\begin{itemize}
\item  Can matter configurations having {\it NLES} be associated with a
particular symmetry of the space time and under what conditions ?

\item  Does the energy-momentum tensor obtained from (\ref{metrica}) through
the Einstein Equations, and satisfying (\ref{global}), describe a physically
reasonable anisotropic fluid ?

\item  Can this fluid represent a bounded gravitational source ?

\item  What types of gravitational collapse scenarios will emerge from
bounded matter configurations with {\it NLES }?
\end{itemize}

The structure of the present paper is guided by the above four questions. In
the next section it is found that a solution of the Einstein equations,
obtained from (\ref{metrica}) restricted by (\ref{condicion}), admits a
Conformal Killing Vector. In Section III it is shown how anisotropic
radiating configurations with {\it NLES} satisfy energy conditions (weak,
strong and dominant) for imperfect fluids. The third question is addressed
in Section IV. There we consider the consequences on the evolution of the
boundary surface when this interior solution is matched to Vaidya exterior
metric. Answers to the last question are explored in sections IV and V by
studying the evolution of models of radiating spheres with a nonlocal
equation of state (either \ref{global} or \ref{global2}). Finally in the
last section our conclusions and results are summarized. If it is not
explicitly stated as ${\rm C}(T)$ (or ${\rm C}(u),$ depending on the tetrad)
we shall assume in this paper ${\rm C}$ as a constant parameter in the above
equation (\ref{condicion}).

\section{Symmetries and {\it NLES }for an anisotropic radiating sphere}

It is well known that there are, at least, two reasons why it is interesting
to start investigating the geometric properties of a particular space time.
The first is a practical one: the formidable difficulties encountered in the
study of the Einstein Field Equations, due to their non-linearity, can be
partially overcame if some symmetry properties of the space-time are
assumed. The second one is more fundamental; symmetries may also provide
important insight and information into the general properties of
self-gravitating matter configurations \cite{Lindblom93}. Guided by the
above motivations we shall show, in this section, that the space times (\ref
{metrica}) having a {\it NLES }(\ref{global} or \ref{global2}) admit a
Conformal Motion ({\it CM}) with ${\rm C}$ considered as a constant
parameter.

In general, {\it CM} is a map $M\rightarrow M$ such that the metric $g$ of
the space time transforms under the rule 
\[
g\rightarrow \tilde{g}={\rm e}^{2\psi }g,\qquad \qquad {\rm with\qquad }\psi
=\psi (x^a). 
\]
This can be expressed as 
\begin{equation}
{\pounds }_{{\bf \xi }}\,{\bf g}_{ab}=%
\mbox{\boldmath $\xi$}
_{a;b}+%
\mbox{\boldmath $\xi$}
_{b;a}=\psi (x^\mu )\,{\bf g}_{ab}\,,  \label{conforme}
\end{equation}
where $%
\mbox{\boldmath $\xi$}
^\mu $ is called a Conformal Killing Vector ({\it CKV}). It is clear that
other important symmetries as homotetic motions or self-similarity ($\pounds
_\xi \,\,g_{ab}=2\sigma \,g_{ab}\ \ $with$\ \sigma =const.$) and isometries (%
$\pounds _\xi \,\,g_{ab}=0$) are particular cases of {\it CM}. This symmetry
imposes an important restriction on the hydrodynamic variables and
consequently it is possible to obtain an equation of state for a space time (%
\ref{metrica}) having a {\it CKV} \cite
{HerreraEtal84,DuggalSharma86,MaartensMasonTsamparlis86,Duggal87}.For the
sake of simplicity and in order to solve the above Conformal Killing
Equations (\ref{conforme}) we shall assume ${\rm C}$ as a constant
parameter. The most general case, ${\rm C}={\rm C}(T)$, will be briefly
considered at the end of the present work.

Additionally, we further restrict our calculations to a vector field $%
\mbox{\boldmath $\xi$}
^\mu $ of the form 
\begin{equation}
\mbox{\boldmath $\xi$}
^\mu =\sigma ( 
\mbox{\scriptsize{$T$}}
,%
\mbox{\scriptsize{$R$}}
)\delta _0^\mu +\lambda ( 
\mbox{\scriptsize{$T$}}
,%
\mbox{\scriptsize{$R$}}
)\delta _1^\mu \;.  \label{vector}
\end{equation}

From (\ref{metrica}) and (\ref{vector}), (\ref{conforme}), with the
condition (\ref{condicion}), we have 
\begin{eqnarray}
\dot{\sigma}+\left[ \beta ^{\prime }-\frac 1R\right] \lambda +{\sigma }\dot{%
\beta} &=&0\,,  \label{cc1} \\
-{\dot{\lambda}}+{\rm C}^2\sigma ^{\prime } &=&0\,,\qquad {\rm and}\qquad
\label{cc2} \\
{\lambda }^{\prime }+\left[ \beta ^{\prime }-\frac 1R\right] \lambda +\sigma 
\dot{\beta} &=&0\,.  \label{cc3}
\end{eqnarray}
where dots and primes denote differentiation with respect to $T$ and $R$,
respectively.

Using (\ref{cc1}) and (\ref{cc3}) we get 
\begin{equation}
-{\ddot{\lambda}}+{\rm C}^2\lambda ^{\prime \prime }=0\,.  \label{cc5}
\end{equation}
A similar expression can be obtained for $\sigma $. The solutions of
equation (\ref{cc5}) and the corresponding equation for $\sigma $ are 
\begin{equation}
\lambda =f({\rm u})+g({\rm v})\,,\qquad {\rm and\qquad }\sigma =m({\rm u})+n(%
{\rm v})\,,  \label{lambda}
\end{equation}
$f({\rm u})$, $g({\rm v})$, $m({\rm u})$ and $n({\rm v})$ being arbitrary
functions of their arguments 
\begin{equation}
{\rm u}={\rm C}T+R\,,\qquad {\rm and\qquad v}={\rm C}T-R\,.
\end{equation}
With equations (\ref{lambda}) in (\ref{cc2}), we obtain

\begin{equation}
\sigma =\frac 1{{\rm C}}\left[ f({\rm u})-g({\rm v})\right] \,.  \label{cc9}
\end{equation}
and, form (\ref{lambda}) and (\ref{cc9}), equation (\ref{cc3}) can be
written as 
\begin{equation}
2\left[ \frac{\partial \beta }{\partial {\rm u}}-\frac 1{{\rm u}-{\rm v}}%
\right] f+\frac{\partial f}{\partial {\rm u}}-2\left[ \frac{\partial \beta }{%
\partial {\rm v}}+\frac 1{{\rm u}-{\rm v}}\right] g-\frac{\partial g}{%
\partial {\rm v}}=0\,.  \label{cc10}
\end{equation}
This can be integrated by demanding that the coefficient of $f({\rm u})$ be
only a function of ${\rm u}$ as well as the coefficient $g({\rm v})$ a
function of ${\rm v}$: 
\begin{equation}
\frac{\partial \alpha ({\rm u})}{\partial {\rm u}}\equiv \frac{\partial
\beta }{\partial {\rm u}}-\frac 1{{\rm u}-{\rm v}}\,,\qquad {\rm and\qquad }%
\frac{\partial \delta ({\rm v})}{\partial {\rm v}}\equiv \frac{\partial
\beta }{\partial {\rm v}}+\frac 1{{\rm u}-{\rm v}}\,.  \label{alfa}
\end{equation}
Thus, we get 
\begin{equation}
\beta ({\rm u},{\rm v})=\ln \left( {\rm u}-{\rm v}\right) +\alpha ({\rm u}%
)+\delta ({\rm v})+{\it C}_1\,,  \label{betauv}
\end{equation}
and equation (\ref{cc10}) may now be written as 
\begin{equation}
2\left( \frac{\partial \alpha }{\partial {\rm u}}\right) f+\frac{\partial f}{%
\partial {\rm u}}=2\left( \frac{\partial \delta }{\partial {\rm v}}\right) g+%
\frac{\partial g}{\partial {\rm v}}={\it k}\,;  \label{cc11}
\end{equation}
where ${\it k}$ and ${\it C}_1$ are constants of integration. This
expression is integrated to give 
\begin{equation}
f({\rm u})={e^{-2\,\alpha }}\left[ {\it k}\int \!{e^{2\,\alpha }}{d{\rm u}}+ 
{\it C}_2\right] \,\qquad {\rm and\qquad }g({\rm v})={e^{-2\,\delta }}\left[ 
{\it k}\int \!{e^{2\,\delta }}{d{\rm v}}+{\it C}_3\right] \,,  \label{fu}
\end{equation}
again, ${\it C}_2$ and ${\it C}_3$ are constants of integration. Then, (\ref
{fu}) implies 
\begin{equation}
\mbox{\boldmath $\xi$}
^\mu =\frac 1{{\rm C}}\left[ f({\rm u})-g({\rm v})\right] \,\delta _0^\mu
+\left[ f({\rm u})+g({\rm v})\right] \,\delta _1^\mu \;\qquad {\rm and\qquad 
}\psi =\frac 1R\left[ f({\rm u})+g({\rm v})\right] \;,  \label{vectorexpl}
\end{equation}
and (\ref{betauv}) may now be written as 
\begin{equation}
\beta (T,R)=\ln (2kR)+\alpha ({\rm C}T+R)+\delta ({\rm C}T-R)+{\it C}_1 \,.
\label{betas2}
\end{equation}
Finally, the metric (\ref{metrica}) reads: 
\begin{equation}
ds^2=R^2\left[ 4k^2e^{2(\alpha +\delta )}\left( {\rm C}\,dT^2-\frac{dR^2}{%
{\rm C}}\right) -d\Omega ^2\right] \,.  \label{bondisch3}
\end{equation}

It should be stressed that the above results have been obtained assuming (%
\ref{alfa}) and ${\rm C}$ as a constant parameter in (\ref{condicion}).

\section{Energy Conditions for Imperfect Fluids}

We have found that in static configurations {\it NLES} are more easily
handled within anisotropic fluids \cite{HernandezNunez98}. Thus, in order to
explore how physically reasonable is the previous non-static fluid with a 
{\it NLES} and having a {\it CKV,} we shall study the energy conditions for
an imperfect anisotropic fluid (unequal stresses, i.e. $P_r\neq P_{\perp }$
and heat flux). Although a pascalian assumption ($P_r=P_{\perp }$) is
supported by solid observational ground, an increasing amount of theoretical
evidence strongly suggests that, for certain density ranges, a variety of
very interesting physical phenomena may take place giving rise to local
anisotropy (see \cite{HerreraSantos97} and references therein).

For these fluids the energy-momentum tensor takes the form 
\begin{equation}
{\bf T}_{\mu \nu }=(\rho +P_{\perp }){\bf u}_\mu {\bf u}_\nu -P_{\perp }{\bf %
g}_{\mu \nu }+(P_r-P_{\perp }){\bf n}_\mu {\bf n}_\nu +{\bf f}_\mu {\bf u}%
_\nu +{\bf f}_\nu {\bf u}_\mu \,,  \label{tuv1}
\end{equation}
where 
\begin{equation}
{\bf u}_\mu =4R^2e^{2(\alpha +\delta )}{\rm C\ }\delta _\mu ^0\,,\qquad {\bf %
n}_\mu =\frac{4R^2}{{\rm C}}e^{2(\alpha +\delta )}\ \delta _\mu ^1\,\qquad 
{\rm and}\qquad {\bf f}_\mu =-q\,{\bf n}_\mu \,,
\end{equation}
with $\rho $, $P_r$, $P_{\perp }$ and $q,$ denoting the energy density, the
radial pressure, the tangential pressure and the heat flow (${\bf f}_\mu 
{\bf f}^\mu =-q^2$), respectively. The fluid four-velocity is represented by 
${\bf u}_\mu $ and ${\bf n}_\mu $ is a unit vector pointing the flux
direction.

The energy conditions are obtained from the eigenvalues of the
energy-momentum tensor (\ref{tuv1}), i.e. they emerge from the roots of the
characteristic equation \cite{KolassisSantosTsoubelis88} 
\begin{equation}
\mid {\bf T}_a^b-\lambda \,%
\mbox{\boldmath $\delta$}
_a^b\mid \,=0\,.  \label{caracter}
\end{equation}
The four eigenvalues take the form 
\begin{equation}
\lambda _0=\frac 12\left[ \rho -P_r+\Delta \right] ,\qquad \,\lambda _1=%
\frac 12\left[ \rho -P_r-\Delta \right] \,,\qquad {\rm and}\qquad \lambda
_2=\lambda _3=-P_{\perp }\,,
\end{equation}
where 
\begin{equation}
\Delta =\left[ (\rho +P_r)^2-4q^2\right] ^{1/2}\,.  \label{deltas}
\end{equation}
With the metric (\ref{bondisch3}), Einstein's field equations can be written
as 
\begin{eqnarray}
8\pi \,{\bf T}_T^T &=&8\pi \,\rho =\frac 1{R^2}+\frac{{\cal Z}}{R^4}\left[
1+2\,R\left( \alpha ^{\prime }-\delta ^{\prime }\right) \right] \,,
\label{ee1} \\
8\pi \,{\bf T}_R^R &=&-8\pi \,P_r=\frac 1{R^2}-\frac{{\cal Z}}{R^4}\left[ 3+{%
\ \ \ 2R}\left( {\alpha }^{\prime }-{\delta }^{\prime }\right) \right] \,,
\label{ee2} \\
8\pi \,{\bf T}_\Theta ^\Theta  &=&-8\pi \,P_{\perp }=\frac{{\cal Z}}{R^4}%
,\qquad {\rm and}\qquad   \label{ee3} \\
8\pi \,{\bf T}_{TR} &=&-8\pi \,{q}=\frac{2{\cal Z}}{R^3}\left( \dot{\alpha}+%
\dot{\delta}\right) \,,  \label{ee4}
\end{eqnarray}
where 
\begin{equation}
{\cal Z}\equiv \frac{{\rm C}}{4e^{2\left( \alpha +\delta \right) }}>0\,.
\label{zeta}
\end{equation}
From two of the above Einstein equations, (\ref{ee1}) and (\ref{ee2}), we
find that the conditions for energy density and radial pressure to be
positive are 
\begin{equation}
\alpha ^{\prime }-\delta ^{\prime }\,\geq \,-\frac R{2{\cal Z}}-\frac 1{2R}%
\,\qquad {\rm and}\qquad \alpha ^{\prime }-\delta ^{\prime }\,\geq \,\frac R{%
2{\cal Z}}-\frac 3{2R}\,.  \label{des2}
\end{equation}
Considering (\ref{ee3}) and (\ref{zeta}) it can  easily be shown that the
tangential pressure, $P_{\perp }$, is negative. Consequently the condition (%
\ref{deltas}) is fulfilled if 
\begin{equation}
2P_{\perp }+\Delta \geq 0\quad \Rightarrow \quad \alpha ^{\prime }-\delta
^{\prime }\,\geq -\,\left( \dot{\alpha}+\dot{\delta}\right) -\frac 1R\,.
\label{des3}
\end{equation}
We then find that the weak energy condition is achieved if: 
\begin{equation}
\rho -P_r+\Delta \geq 0\,\quad \Rightarrow \quad \Delta \geq \,\frac{2{\cal Z%
}}{R^4}-\frac 2{R^2}\,,  \label{debil}
\end{equation}
the dominant energy condition is held if 
\begin{equation}
\rho -P_r\geq 0\Rightarrow \frac{{\cal Z}\,}{R^2}\leq \,1\,\qquad {\rm and}%
\qquad \rho -P_r-2P_{\perp }+\Delta \geq 0\,\Rightarrow \Delta \geq \,-\frac 
2{R^2}\,,  \label{domi1}
\end{equation}
and the strong energy condition is accomplished if 
\begin{equation}
2P_{\perp }+\Delta \geq 0\Rightarrow \Delta \geq \,\frac{2{\cal Z}}{R^4}\,.
\label{fuerte}
\end{equation}

Next, we are going to show that it is possible to fulfill simultaneously all
of the above energy conditions. From the definition for ${\cal Z}$ and (\ref
{domi1}) it is now seen that 
\begin{equation}
0<\frac{{\cal Z}\,}{R^2}\leq \,1\,.  \label{des4}
\end{equation}
Therefore, inequalities (\ref{debil}), (\ref{domi1}) and (\ref{fuerte})
yield 
\begin{equation}
\frac{\Delta R^2}2\geq \frac{{\cal Z}\,}{R^2}\,\Rightarrow e^{2\left( \alpha
+\delta \right) }\geq \frac{{\rm C}}{2\Delta R^4}.  \label{fina0}
\end{equation}
Now, considering (\ref{des4}), (\ref{des2}) and (\ref{des3}) we have 
\begin{equation}
\alpha ^{\prime }-\delta ^{\prime }\geq -\frac 1R\,\qquad {\rm and}\qquad 
\dot{\alpha}+\dot{\delta}\geq 0\,.  \label{fina2}
\end{equation}
By using equation (\ref{betas2}) we can also see that 
\begin{equation}
\beta ^{\prime }=\frac 1R+\alpha ^{\prime }-\delta ^{\prime }\,,
\end{equation}
which, with the help of (\ref{fina2}) becomes 
\begin{equation}
\beta ^{\prime }\geq 0\,.
\end{equation}
This is a well known restriction on the metric function $\beta $ \cite
{Bondi64}. Therefore, it is clear that matter configurations satisfying a 
{\it NLES} and having a {\it CKV} can represent reasonable fluids in General
Relativity. Next section will be devoted to study the consequences of
matching this type of fluids to an exterior solution.

\section{Junction Conditions}

In order to address the third question, we shall demand that the line
element (\ref{bondisch3}) join the Vaidya exterior metric at the boundary of
the configuration. We shall also recall that two regions of a space-time
match across a separating hypersurface $S$ if the first and the second
fundamental forms are continuous across this boundary surface $S$ \cite
{BonnorOliveiraSantos89}.

The Vaidya line element outside the source can be written as 
\begin{equation}
ds_{\left( +\right) }^2=Hdu^2+2dudr-r^2d\Omega ^2\,,\qquad \qquad {\rm with}%
\qquad H=1-\frac{2M\left( u\right) }r\,,
\end{equation}
where the subscript $\left( +\right) $ refers to the exterior region of the
space time. In these coordinates the boundary equation takes the form: $%
r=r_s(u)$, where the subscript $s$ reminds that the corresponding function
has been evaluated at boundary of the configuration. Then, the induced
metric on the boundary surface, from the outside, is 
\begin{equation}
\left( ds^2\right) _s^{\left( +\right) }=\left[ H_s+2\frac{dr_s}{du}\right]
du^2-r_s^2d\Omega ^2\,,  \label{vaidya2}
\end{equation}
and the line element on the boundary, from the inside, reads 
\begin{equation}
\left( ds^2\right) _s^{\left( -\right) }=4R_s^2e^{2(\alpha _s+\delta
_s)}\left[ {\rm C}-\frac 1{{\rm C}}\left( \frac{dR_s}{dT}\right) ^2\right]
dT^2-R_s^2d\Omega ^2\,,  \label{bondisch3s}
\end{equation}
where we have used the fact that the equation of the boundary in the
coordinates $(T,R,\Theta ,\Phi )$ reads: $R=R_s(T)$. Demanding continuity of
the first fundamental form, we get at once 
\begin{eqnarray}
2R_se^{(\alpha _s+\delta _s)}\left[ {\rm C}-\frac 1{{\rm C}}\left( \frac{dR_s%
}{dT}\right) ^2\right] ^{1/2}dT &=&\left[ H_s+2\frac{dr_s}{du}\right]
^{1/2}du\,,\qquad {\rm and}\qquad   \label{funda1} \\
R_s\left( T\right)  &=&r_s\left( u\right) \,.  \label{funda2}
\end{eqnarray}
It can be shown \cite{Herrera96} that (\ref{funda1}) is equivalent to 
\begin{equation}
e^{2(\alpha _s+\delta _s)}=\frac 1{4R{_s^2}}\;.  \label{alfadelta1}
\end{equation}
Also, a straighforward calculation shows that the continuity of the second
fundamental form across the boundary surface is equivalent to the condition 
\cite{BonnorOliveiraSantos89} 
\begin{equation}
\left[ q\right] _s=\left[ P_r\right] _s\,.  \label{acopla3}
\end{equation}
Using (\ref{ee2}) and (\ref{ee4}), the equation (\ref{acopla3}) gives the
evolution of the boundary surface: 
\begin{equation}
R_s=\frac \kappa 2\frac 1{\dot{\alpha _s}+\dot{\delta _s}+\alpha _s^{\prime
}-\delta _s^{\prime }}\;,  \label{alfadelta2}
\end{equation}
where the constant $\kappa $ is : 
\begin{equation}
\kappa \equiv \frac{1-3{\rm C}}{{\rm C}}\;.  \label{constka}
\end{equation}
Now, substituting equation (\ref{alfadelta2}) in (\ref{alfadelta1}), we
obtain 
\begin{equation}
\dot{\alpha _s}+\dot{\delta _s}+\alpha _s^{\prime }-\delta _s^{\prime
}=\kappa \,e^{\alpha _s+\delta _s}\;.  \label{alfadelta3}
\end{equation}
It is useful to write equation (\ref{alfadelta3}) in terms of the variables $%
{\rm u}$ and ${\rm v}$ 
\begin{equation}
\left( \frac{\partial {\alpha _s}}{\partial {\rm u}}+\frac{\partial {\delta
_s}}{\partial {\rm v}}\right) \left( {\rm C}+1\right) =\kappa \,e^{\left(
\alpha _s\left( {\rm u}\right) +\delta _s\left( {\rm v}\right) \right) }\;,
\label{alfadelta4}
\end{equation}
which can be easily integrated for constant $\alpha _s$. Thus, assuming $%
\alpha _s=$ $K=const$, we get 
\begin{equation}
\delta _s=-K-\ln \left[ -\frac{\kappa \left( {\rm v}-\left( {\rm C}+1\right)
C_2\right) }{{\rm C}+1}\right] \;.  \label{alfadelta5}
\end{equation}
Next, considering $K$ and $\delta _s$ in equation (\ref{alfadelta1}), the
evolution of the boundary surface can be written as 
\begin{equation}
R_s=\frac{\kappa \left[ {\rm C}T-\left( {\rm C}+1\right) \,C_2\right] }{%
\kappa \pm 2\left( {\rm C}+1\right) }\;,  \label{radiomasmenos}
\end{equation}
where $C_2$ is a constant of integration. It is clear that there are two
possible evolutions for the bounding surface and in both of them $R_s$ is a
linear function of the time coordinate, namely

\begin{eqnarray}
R_s^{+} &=&\frac{\left( 1-3{\rm C}\right) {\rm C}}{2{\rm C}^2-{\rm C}+1}\ T-%
\frac{\left( 1-3{\rm C}\right) \left( {\rm C}+1\right) C_2^{+}}{2{\rm C}^2-%
{\rm C}+1}\ ,\qquad {\rm and}\qquad   \label{radio1} \\
R_s^{-} &=&\frac{\left( 1-3{\rm C}\right) {\rm C}}{-2{\rm C}^2-5{\rm C}+1}\
T-\frac{\left( 1-3{\rm C}\right) \left( {\rm C}+1\right) C_2^{-}}{-2{\rm C}%
^2-5{\rm C}+1}\ .  \label{radio2}
\end{eqnarray}
Notice that expression (\ref{constka}) for the constant $\kappa $ has been
used and we have denoted $C_2^{+}$ and $C_2^{-}$ the corresponding constants
for the two equations emerging from (\ref{radiomasmenos}).

Let us consider each of the above solutions separately. First, it is clear
that in (\ref{radio1}), due to 
\begin{equation}
2{\rm C}^2-{\rm C}+1>0\,\,\,\qquad \forall \,\,\,{\rm C\ }\in {\rm \ }\left(
0,1\right) \ ,  \label{R1denomin}
\end{equation}
expanding configurations can be obtained if ${\rm C}\,\,\in \,\,\left( 0\;,\;%
\frac 13\right) $ and collapsing ones for ${\rm C~}\in \left( ~\frac 13%
\;,~1\right) .$ Additionally, because$\;$ $R_s^{\pm }>0$ $\quad \forall \
T\in \left( 0,\infty \right) $, several values of the possible initial
configurations ($M_0$ and $R_s^{\pm }$) have to be excluded. Concerning the
first case, $R_s^{+}$, the allowed values for the constant ${\rm C}\,$are

\begin{eqnarray}
{\rm C}\,\, &\in &\,\,\left( 0\;,\;\frac 13\right) \,\,\,\qquad \Rightarrow
\ {\rm Expanding\ Configurations,}\qquad {\rm and}\qquad \\
{\rm C}\,\, &\in &\,\,\left( \frac 13\;,\;1\right) \qquad \Rightarrow \ \ 
{\rm Contracting\ Configurations.}  \label{contracting1}
\end{eqnarray}
In the second case (equation (\ref{radio2})), the permitted values can be
expressed as 
\begin{eqnarray}
{\rm C}\,\, &\in &\,\,\left( 0\;,\;\frac 14\left( \sqrt{33}-5\right) \right)
\cup \,\,\,\left( \frac 13\;,\;1\right) \qquad \Rightarrow \ {\rm Expanding\
Configurations,}\qquad {\rm and}\qquad \\
{\rm C}\,\, &\in &\,\,\left( \frac 14\left( \sqrt{33}-5\right) \,,\,\frac 13%
\right) \;\qquad \Rightarrow \ \ {\rm Contracting\ Configurations.}
\label{contracting2}
\end{eqnarray}
It is clear that $C_2^{+}$ and $C_2^{-}$can be solved from (\ref{radio1}),
and (\ref{radio2}), namely 
\begin{eqnarray}
C_2^{+} &=&-\frac{R_{so}^{+}\left( 2\,{\rm C}^2-{\rm C}+1\right) }{\left(
1-3\,{\rm C}\right) \left( {\rm C}+1\right) }\ ,\qquad {\rm and}\qquad \\
C_2^{-} &=&\frac{R_{so}^{-}\left( 2\,{\rm C}^2+5\,{\rm C}-1\right) }{\left(
1-3\,{\rm C}\right) \left( {\rm C}+1\right) }\ ,
\end{eqnarray}
with $R_{so}^{+}$ and $R_{so}^{-}$ the initial values for $R_s^{+}$ and $%
R_s^{-}$, respectively. For the present work we shall only consider those
values of ${\rm C}$ that induce collapsing configurations.

For this particular matter distribution, starting with an initial radius, $%
R_{so}^{\pm },$ and mass, $M_0$, we shall define the{\it \ Total Evolution
Time}, $Tev$, as the time it takes to radiate away its total initial mass $%
M_0$. Easily, we get the expressions for the $Tev$ corresponding to (\ref
{radio1}), and (\ref{radio2}), concerning the values of the constant
parameter ${\rm C}$ in (\ref{contracting1}) and (\ref{contracting2}),
respectively. Thus, they can be written as 
\begin{eqnarray}
Tev_I &=&\frac{R_{so}^{+}\left( 1-3\,%
{\displaystyle {M_0 \over R_{so}^{+}}}
\,+\left( 
{\displaystyle {2M_0 \over R_{so}^{+}}}
\right) ^2\right) }{\left( 1-3\,%
{\displaystyle {M_0 \over R_{so}^{+}}}
\,\right) \left( 1-2%
{\displaystyle {M_0 \over R_{so}^{+}}}
\,\right) }\ ,\qquad {\rm and}\qquad   \label{tiempo1} \\
Tev_{II} &=&-\frac{R_{so}^{-}\left( 3\,-9\,%
{\displaystyle {M_0 \over R_{so}^{-}}}
+\,\left( 
{\displaystyle {2M_0 \over R_{so}^{-}}}
\right) ^2\right) }{\left( 1-3\,%
{\displaystyle {M_0 \over R_{so}^{-}}}
\,\right) \left( 1-2%
{\displaystyle {M_0 \over R_{so}^{-}}}
\,\right) }\ .  \label{tiempo2}
\end{eqnarray}
It is clear that given the initial values for the mass and radius
(therefore, the parameter ${\rm C=}$ $\frac{R_{so}^{\pm }}{M_0}$ ), $Tev$
for both evolutions is fully determined. It is also worth mentioning the
particular astrophysical scenario that the evolution of these matter
configurations could simulate. Both of them (either (\ref{radio1}), or (\ref
{radio2})) start at a particular value of the parameter ${\rm C}$,
completely radiate away their initial mass and disappear leaving no remnant.
In this sense this scenario resembles Type I Supernova but, in the present
case, it begins from a neutron star progenitor (characteristic values for
the radius, $R_{so}^{\pm }$ and the mass $M_0$) and gravitation plays an
important role. In the standard Type I Supernova picture there is a white
dwarf progenitor and gravitation is not longer the main cause of the
emission processes.

In Figure 1 the total evolution times $Tev_I$ and $Tev_{II}$ are sketched as
a function of the initial mass. In both cases we have assumed 10 Kms. as the
value for the initial radius. This value can be considered as typical for
neutron stars \cite{Demianski85,ShapiroTeukolsky83}. It can be appreciated
from this figure that as the initial mass $M_0$ rises, the total time, $%
Tev_I,$ increases asymptotically up to the value $M_0=\frac{R_{so}^{-}}3\,$.
The other branch $Tev_{II}$ decreases coming from this same asymptotic
maximum for $M_0$. These two branches of the evolution time may represent
different astrophysical phenomena. In first case, $M_0<\frac{R_{so}^{-}}3$ ,
it is apparent that the more massive the configuration is, the more time it
takes to radiate the initial mass. In the second branch, $M_0>\frac{%
R_{so}^{+}}3,$ more massive spheres lead to more violent collapses. The
value of $M_0=\frac{R_{so}^{-}}3$ emerges as a critical point of stability
for a configuration with a {\it NLES}, having a {\it CKV} and matched to a
Vaidya exterior solution. Any perturbation accreting over (or expelling mass
from) this type of relativistic sphere will initiate processes which radiate
away a significant amount of energy.

In the next section we shall consider the hydrodynamic consequences of
configuration satisfying a {\it NLES} for both cases: ${\rm C=}const.$ and $%
{\rm C=C}(T)${\it .}

\section{Collapsing Radiating Spheres\label{spheres}}

Finally, to determine if anisotropic fluid spheres with a {\it NLES} can
conform a feasible matter candidate to stellar models, we shall study
collapsing scenarios for radiating configuration of these type of materials.
In what follows, radiating collapsing models will be matched to the Vaidya
exterior metric and energy density will always be positive and larger than
pressure everywhere within the fluid distribution. Finally, the fluid
velocity, as measured by the locally Minkowskian observer, will always be
less than one.

Besides these ``regularity'' conditions, the modeling emerges from a
heuristic assumption relating density, pressure and radial matter velocity.
This {\it ansatz}, guided by solid physical principles, allows the
generation of radiating solutions from a known static ``seed'' solutions and
reduces the problem of solving Einstein Equations to a numerical integration
of a system of ordinary differential equations for quantities evaluated at
the surfaces (shocks and/or boundaries). The rationale behind this {\it %
ansatz} can be grasped in terms of the characteristic times for different
processes involved in a collapse scenario. If the hydrostatic time scale, $%
{\cal T}_{HYDR}$, which is of the order $\sim 1/\sqrt{G\rho }$ (where $G$ is
the gravitational constant and $\rho $ denotes the mean density), is much
smaller than the {\it Kelvin-Helmholtz} time scale (${\cal T}_{KH}$), then
in a first approximation, the inertial terms in the equation of motion can
be ignored \cite{KippenhahnWeigert90}.

\subsection{The metric and the matter}

The method we use to build these models is a general strategy presented
several years ago by L. Herrera, J. Jim\'{e}nez and G. Ruggeri (HJR). Only a
very brief description of this method is given here. We refer the reader to 
\cite{HerreraJimenezRuggeri80} and \cite{HerreraNunez90} for details. In
this method a nonstatic spherically symmetric distribution of matter is
assumed. The metric representing this distribution, in Bondi radiation
coordinates \cite{BondiVandenburgMetzner62}, takes the form 
\begin{equation}
ds^2=e^{4\beta \left( u,r\right) }h\left( u,r\right) \ du^2+2e^{2\beta
\left( u,r\right) }\ du\,dr-r^2(d\vartheta ^2+\,\sin ^2\phi \,d\phi
^2)\,,\qquad {\rm with\quad }h\equiv 1-\frac{2{m}\left( u,r\right) }r.
\label{bondi}
\end{equation}
Here $u=x^0$ is a time like coordinate, $r=x^1$ is the null coordinate and $%
\theta =x^2$ and $\phi =x^3$ are the usual angle coordinates. The $u$
-coordinate is the retarded time in flat space-time and, therefore, $u$
-constant surfaces are null cones open to the future. The function ${m}%
\left( u,r\right) $ is the generalization, inside of the distribution, of
the ``mass aspect'' defined by Bondi \cite{BondiVandenburgMetzner62} which
in the static limit coincides with the Schwarzschild mass. In addition we
consider an anisotropic fluid sphere composed by a material medium plus
radiation. For a observer with a radial velocity $-\omega $ with respect to
the fluid, the energy-momentum tensor can be written as \cite
{HerreraNunez90,Martinez96} 
\begin{equation}
{\bf T}_{\mu \nu }=(\rho +P_{\bot }){\bf u}_\mu {\bf u}_\nu -P_{\bot }{\bf g}%
_{\mu \nu }+(P_r-P_{\bot }){\bf n}_\mu {\bf n}_\nu +{\bf f}_\mu {\bf u}_\nu +%
{\bf f}_\nu {\bf u}_\mu \,,  \label{tuv6}
\end{equation}
with 
\begin{equation}
{\bf u}_\mu =\frac 1{(1-\omega ^2)^{\frac 12}}\left[ h^{\frac 12}e^{2\beta
}\delta _\mu ^0+\frac{1-\omega }{h^{\frac 12}}\delta _\mu ^1\right] \,,\quad 
{\bf n}_\mu =\frac 1{(1-\omega ^2)^{\frac 12}}\left[ -\omega \delta _\mu ^0+%
\frac{1-\omega }{h^{\frac 12}}\delta _\mu ^1\right] \,,\quad {\bf f}_\mu
=-q\,{\bf n}_\mu \,,
\end{equation}
and where the physical variables, as measured by a local {\it minkowskian }
observer, are represented by the energy density, $\rho $; the radial and
tangential pressure, $P_r,$ $P_{\bot }$, respectively; the radiation energy
flux density ${\bf f}_\mu $ and fluid radial velocity, $-\omega $. The
co-moving {\it minkowskian} observer coincides with the{\it \ Lagrangean
frame} (the {\it proper frame}) where the interaction between radiation and
matter can be easily handled \cite{MihalasMihalas84}. Notice that the
radiation is treated in the diffusive regime, therefore it is considered to
have a mean free path much smaller than the characteristic length of the
system. Within this regime, radiation is locally isotropic and we have 
\begin{equation}
\left. 
\begin{array}{c}
\rho _R=3{\cal P} \\ 
{\cal F}=q
\end{array}
\right\} \ \Longrightarrow P_r=\hat{P}+{\cal P}\,.
\end{equation}
Where, $\hat{P}$, describes the fluid radial pressure and the radiation
contribution to the energy density, energy flux density and radial pressure
are represented by $\rho _R$, ${\cal F}$ and ${\cal P}$, respectively. It is
clear that in this radiation limit the total radial pressure, $P_r$,
encompasses the hydrodynamic and the radiation contribution to the pressure
(see \cite{AguirreHernandezNunez94} for details).

\subsection{The Einstein Field Equations and the Matching}

Inside the matter distribution the Einstein field equations can be written
as 
\begin{eqnarray}
\frac{\rho +2\omega q+\omega ^2P_r}{1-\omega ^2}-\frac 1{4\pi r^2}\left( -%
\frac{{\dot m}}{h\, e^{2\beta}}+{m}^{\prime}\right) &=&0\,,  \label{ebb1} \\
\frac{\rho -\omega P_r-(1-\omega )q}{1+\omega }-\frac 1{4\pi r^2}{m}%
^{\prime} &=&0\,,  \label{ebb2} \\
\frac{(\rho +P_r-2q)(1-\omega )}{1+\omega }-\frac h{2\pi r}\beta^{\prime}
&=&0\,,\qquad {\rm and}\qquad  \label{ebb3} \\
P_{\perp }+\frac{\dot{\beta}^{\prime}}{4\pi e^{2\beta }}-\frac h{8\pi }
\left( 2\beta^{\prime \prime}+4\left( \beta^{\prime} \right) ^2- \frac{%
\beta^{\prime}}r\right) -\frac{3\beta^{\prime}(1-2{m}^{\prime})- {m}^{\prime
\prime}}{8\pi r} &=&0\,.  \label{ebb4}
\end{eqnarray}
As was stressed by Bondi \cite{Bondi64} it is possible to solve
algebraically the physical variables of equations (\ref{ebb1}) - (\ref{ebb4}%
) if we have both ${m}\left( u,r\right) $ and $\beta \left( u,r\right) $
(and their derivatives) plus another equation that relates the tangential
and radial pressure. This method will be clarified by an example in the next
section.

From the field equations (\ref{ebb2}) and (\ref{ebb3}) it is clear that they
can be (formally) integrated yielding 
\begin{eqnarray}
{m} &=&4\pi \int_0^r\frac{\rho -\omega P_r-(1-\omega )q}{1+\omega }%
r^2dr=4\pi \int_0^rr^2\tilde{\rho}dr\,,\qquad {\rm and}\qquad  \label{eme} \\
\beta &=&2\pi \int_a^r\frac{(\rho +P_r-2q)(1-\omega )}{1+\omega }\frac rh%
dr=2\pi \int_a^r\left( \tilde{\rho}+\tilde{P}\right) \frac rhdr\,,
\label{beta}
\end{eqnarray}
where 
\begin{equation}
\tilde{\rho}\equiv \frac{\rho -\omega P_r-(1-\omega )q}{1+\omega }\,,\qquad 
{\rm and\qquad }\tilde{P}\equiv \frac{-\omega \rho +P_r-(1-\omega )q}{%
1+\omega }\,.  \label{denef}
\end{equation}
These auxiliary functions $\tilde{\rho}$ and $\tilde{P}$ must coincide with
the energy density and the radial pressure in the static limit.

Matching the Vaidya metric to the Bondi metric (\ref{bondi}) at the surface
of the fluid distribution, $r=a$, implies $\beta _a=0$ with the continuity
of the mass function ${m}\left( u,r\right) $, i.e. the continuity of the
first fundamental form. The continuity of the second fundamental form leads
to (again, see \cite{AguirreHernandezNunez94} for details) 
\begin{equation}
\tilde{P}_a=-\omega _a\,\tilde{\rho}_a\;.  \label{condisup}
\end{equation}
The subscript $a$ indicates that the corresponding quantity is evaluated at
the boundary surface.

\subsection{Surface equations}

As we have mentioned, the crucial point of the HJR method is the assumption
that the radial dependence of $\tilde{\rho}$ and $\tilde{P}$ can be borrowed
from a ``seed'' static solution and, therefore the metric functions ${m}%
\left( u,r\right) $ and $\beta \left( u,r\right) $ can be determined from
eq. (\ref{eme}) and (\ref{beta}), up to some functions of the time-like
coordinate $u,$ which in turn are computed by integrating a system of
ordinary differential equation coming from the boundary conditions. This
assumption represents a correction to the first approximation in the radial
velocity (i.e., quasi-stationary approximation) and is expected to yield
good results whenever ${\cal T}_{KH}\gg {\cal T}_{HYDR}$. Fortunately
enough, this condition is fulfilled for almost all kind of stellar objects.
For example, in the case of the Sun we get ${\cal T}_{KH}\sim 107$ years,
whereas ${\cal T}_{HYDR}\sim 27$ minutes. Also, the {\it \ Kelvin-Helmholtz }%
phase of the birth of a neutron star last for about tens of seconds, whereas
for a neutron star of one solar mass and a ten kilometer radius, we obtain $%
{\cal T}_{HYDR}\sim 10^{-4}sec$.

It is clear that with the metric functions completely determined, the
physical variables ($\rho $, $P_r,$ $P_{\bot }$ ${\bf f}_\mu $ , and $\omega 
$) are algebraically solved from the field equations (\ref{ebb1}) through (%
\ref{ebb4}).

Scaling the radius $a$, the total mass $m_a$ and the timelike coordinate $u$
by the total initial, mass $m_a\left( 0\right) $, i.e. 
\begin{equation}
A \left(u\right) \equiv \frac{a(u)}{m_a\left(0\right) },\,\,\,\,\,\,
M\left(u\right) \equiv \frac{m_a(u)}{m_a\left(0\right) }, \qquad {\rm and}
\,\,\,\,u \equiv \frac u{m_a\left( 0\right) }\,,  \label{adimen1}
\end{equation}
and defining 
\begin{equation}
F \left(u\right) \equiv 1-\frac{2 M\left(u\right) } {A \left(u\right)},
\qquad {\rm and\qquad } \,\,\,\,\,\,\,\, \Omega \left(u\right) \equiv \frac 1
{1-\omega _a}\ \,.  \label{adimen2}
\end{equation}

The first surface equation comes from the definition of the velocity in
radiation coordinates, as 
\begin{equation}
\dot{A}=F\left( \Omega -1\right) \,.  \label{apunto}
\end{equation}
The second surface equation arises from the luminosity evaluated at the
surface which can be written as 
\begin{equation}
L=-\dot{M}=4\pi A^2q_a\left( 2\Omega -1\right) F\,.  \label{mpunto}
\end{equation}
Using equation (\ref{apunto}) and the definitions (\ref{adimen2}), we can
re-state equation (\ref{mpunto}) as 
\begin{equation}
\dot{F}=\frac{2L+F\left( 1-F\right) \left( \Omega -1\right) }A\,.
\label{fpunto}
\end{equation}

Finally, some straightforward manipulations coming from the field equations (%
\ref{ebb2}), (\ref{ebb2}), (\ref{ebb3}) and (\ref{ebb4}), lead to 
\begin{equation}
\left[ -\frac d{du}\left( \frac{\tilde{\rho}+\tilde{P}}h\right) \right] +%
\tilde{R}-\frac 2r\left( P_r-\tilde{P}\right) =0\,,  \label{tov1}
\end{equation}
where 
\begin{equation}
\tilde{R}=\tilde{P}^{\prime }+\left( \frac{\tilde{\rho}+\tilde{P}}h\right)
\left( 4\pi r\tilde{P}+\frac m{r^2}\right) -\frac 2r\left( P_{\bot
}-P_r\right) \,.
\end{equation}
Notice that (\ref{tov1}) corresponds to a generalization of the
Tolman-Oppenheimer-Volkov equation for any dynamic radiative situation.
Evaluating it on the boundary surface $r=a_{-}$ and using the definitions (%
\ref{adimen2}), the third surface equation can be obtained, i.e. 
\begin{equation}
\frac{\dot{F}}F+\frac{\dot{\Omega}}\Omega -\frac{\dot{\tilde{\rho}_a}}{%
\tilde{\rho}_a}+\frac{F\Omega ^2}{\tilde{\rho}_a}\tilde{R}_a-\frac{2F\Omega }%
A\frac{P_{ra}}{\tilde{\rho}_a}-G=0\,,  \label{tercera}
\end{equation}
where 
\begin{equation}
G=\left( 1-\Omega \right) \left[ \frac{4\pi A\left( 3\Omega -1\right) }\Omega
\tilde{\rho}_a-\frac{3+F}{2A}+F\Omega \frac{\tilde{\rho}_a^{\prime }}{\tilde{%
\rho}_a}+\frac{2F\Omega }{A\tilde{\rho}_a}\left( P_{\bot }-P_r\right)
_a\right] \,.
\end{equation}

Equations (\ref{apunto}), (\ref{fpunto}) and (\ref{tercera}) constitute the 
{\it System of Surface Equations }({\it SSE}). This system can be integrated
for any given radial dependence of the effective variables $\tilde{\rho}$
and $\tilde{P}$. In the context HJR with anisotropy a general equation has
to be provided in order to relate the tangential and the radial pressure 
\cite{CosenzaEtal82,HerreraNunez90,Martinez96}. 
\begin{equation}
P_{\perp }\left( u,r\right) -P_r\left( u,r\right) =\frac 12\left[ r\tilde{P}%
^{\prime }+\frac{\tilde{\rho}+\tilde{P}}{1-\frac{2m}r}\left( \frac mr+4\pi
r^2\tilde{P}\right) \right] \,.  \label{mianis}
\end{equation}
It is clear that if the anisotropic equation of state (\ref{mianis}) is
given, we have three differential equations to get four unknown functions,
i.e. the exterior radius $A(u),$ the gravitational potential at the surface, 
$F(u),$ the radial velocity, $\Omega (u),$ and the luminosity profile, $L(u).
$ Since the only observable quantity entering a ``real'' gravitational
collapse is the luminosity, it seems reasonable to provide such a profile as
an input for our modelling.

\subsection{The HJR formalism and {\it NLES}}

At this point we shall verify that, in the HJR formalism, models having (\ref
{condicion}), satisfy a nonlocal equation (\ref{global} or \ref{global2})
within the distribution. Taking derivatives of (\ref{condicion}) with
respect to $r$, we get 
\begin{equation}
-\frac{m^{\prime }}r+\frac m{r^2}=-\beta ^{\prime }e^{-2\beta }\,{\rm C}%
(u)\,,  \label{ecua1}
\end{equation}
and using (\ref{condicion}), (\ref{eme}) and (\ref{beta}) with (\ref{ecua1})
we obtain 
\begin{equation}
m=2\pi r^3\left( \tilde{\rho}-\tilde{P}\right) \,.  \label{ecua2}
\end{equation}
On the other hand, taking derivatives of (\ref{ecua2}) with respect to $r$
and comparing it with (\ref{eme}) it follows that 
\begin{equation}
\tilde{\rho}-3\tilde{P}+r\left( \tilde{\rho}^{\prime }-\tilde{P}^{\prime
}\right) =0\,.  \label{ecua4}
\end{equation}
As expected, this expression emerges from the condition (\ref{condicion})
and in terms of the field equations it reads 
\begin{equation}
{\bf G}_u^u+3{\bf G}_r^r+r\left[ \left( {\bf G}_u^u\right) ^{\prime }+\left( 
{\bf G}_r^r\right) ^{\prime }\right] =0\,.  \label{ecua55}
\end{equation}
It should be noticed that from equation (\ref{ecua4}) two expressions for 
{\it NLES} can be obtained. They can be written as 
\begin{eqnarray}
\tilde{\rho}(r,u) &=&\tilde{P}(r,u)+\frac 2r\,{\int }_0^r\tilde{P}\,({\bar{r}%
},u){\rm d}{\bar{r}\ +\ }\frac{{\cal H}(u)}r\ ,\qquad {\rm or}\qquad 
\label{globalefect} \\
\tilde{P}(r,u) &=&\tilde{\rho}(r,u)-\frac 2{r^3}\int_0^r\bar{r}^2\tilde{\rho}%
(\bar{r},u){\rm d}\bar{r}\ +\frac{{\cal B}(u)}{r^3}\ .  \label{globalefect2}
\end{eqnarray}
In the present case, equations (\ref{globalefect}) and (\ref{globalefect2})
are (\ref{global}) and (\ref{global2}), respectively, expressed in terms of
the effective variables defined in (\ref{denef}). From this new aspect of
the {\it NLES}, it is clear that the contribution to the pressure and the
energy density coming from the radiation energy flux, as well as the effects
of the fluid velocity, are taken into account in the nonlocal character of
the equation of state. Moreover, equation (\ref{sigmatmunu}), in terms of
the effective variables, becomes 
\begin{equation}
\tilde{P}(r,u)=\frac 13\tilde{\rho}(r,u)-2\left( \frac 13\tilde{\rho}(r,u)-%
\frac 13\left\langle \tilde{\rho}(r,u)\right\rangle \right) +\frac{{\cal B}%
(u)}{r^3}.  \label{sigmaefect}
\end{equation}
Clearly, it is the dynamic generalization of equation (\ref{densaver}), with
the corresponding interpretation of $\tilde{P}(r,u)=\frac 13\tilde{\rho}%
(r,u),$ as the ``effective equation of state'' which may be affected by the
standard deviation 
\begin{equation}
\sigma _{\frac 13\tilde{\rho}(r,u)}=\left( \frac 13\tilde{\rho}(r,u)-\frac 13%
\left\langle \tilde{\rho}(r,u)\right\rangle \right) ,
\end{equation}
coming from the contribution of the nonlocal term $\left\langle \tilde{\rho}%
(r,u)\right\rangle $.

There is an other important consequence arising from (\ref{condicion}) in
terms of the surface variables in the HJR\ formalism. First, comparing the
condition (\ref{condicion}) with the definition for $F\left( u\right) $, it
follows 
\begin{equation}
h_a=1-2\frac MA=F\left( u\right) ={\rm C}(u)\,.  \label{condsup}
\end{equation}
Notice that ${\rm C}(u)\,$ is related to the gravitational potential at the
surface. Next, evaluating (\ref{ecua2}) at the surface and using (\ref
{condisup}) we get 
\begin{equation}
m_a=2\pi a^3\left( 1+\omega _a\right) \tilde{\rho}_a\,.  \label{ecua3}
\end{equation}
This equation can be written as a function of the dimensionless variables (%
\ref{adimen1}) as 
\begin{equation}
1-{\rm C}\left( u\right) =\frac{4\pi A\left( u\right) ^2}{\Omega \left(
u\right) }\left( 2\Omega \left( u\right) -1\right) \tilde{\rho}_a\left(
u\right) \,,
\end{equation}
therefore, the ``velocity'' of the surface $\Omega \left( u\right) $ can be
obtained 
\begin{equation}
\Omega \left( u\right) =\frac{4\pi A\left( u\right) ^2\tilde{\rho}_a\left(
u\right) }{8\pi A\left( u\right) ^2\tilde{\rho}_a\left( u\right) +{\rm C}%
\left( u\right) -1}\,.  \label{omega}
\end{equation}
It is clear that it relates the velocity and the effective density at the
surface.

Now we are going to provide several ``seed'' solutions to explore reasonable
collapsing scenarios compatible with a {\it NLES} in the HJR formalism.

\subsection{Collapsing Scenarios}

The first example considered as a starting static equation of state is the 
{\it \ Schwarzschild} solution. This model represents a generalization of a
radiating incompressible anisotropic fluid of homogeneous density. The
effective density can be written as 
\begin{equation}
\tilde{\rho}_{sh}=\frac 1{8\pi }f\left( u\right) \,.
\end{equation}
With equation (\ref{eme}) evaluated at the surface we find $f\left( u\right) 
$ and using (\ref{omega}) we obtain that the fluid velocity at the surface
is constant and takes the value $\omega _a=-\frac 13c$ . 

The second case of study is the anisotropic {\it Tolman-VI-like} model.
Again, in the static limit this solution is not deprived of a physical
meaning. The static {\it Tolman VI }solution approaches that of a highly
relativistic Fermi Gas and, therefore, one with the corresponding adiabatic
exponent of 4/3. In this case we have

\begin{equation}
\tilde{\rho}_{T_{VI}}=\frac 1{8\pi r^2}g\left( u\right) \,.
\end{equation}
Again equation (\ref{eme}), can be evaluated at the surface, with the
restriction (\ref{omega}), leading to a constant velocity at the surface: $%
\omega _a=c.$

Although the above equations of state have been successfully worked out
within HJR formalism, in our study with a {\it NLES} they lead to very
restrictive matter configurations where the velocity of the surface is held
constant during the collapse. It is worth mentioning that the above results
for the {\it Schwarzschild-like} and {\it Tolman-VI-like} models are valid
for the general case ${\rm C=C}(u).$

Next, we use as a ``seed'' another static equation of state which represents
a richer description for ultracompact nuclear matter. We shall explore the
consequences of adopting a static solution proposed by M. K. Gokhroo and A.
L. Mehra in 1994 \cite{GokhrooMehra94}. This solution corresponds to an
anisotropic fluid with variable density. It has been used in the context of
the HJR method to study several interesting properties of an anisotropic
fluid \cite{Martinez96} describing the Kelvin-Helmoltz phase in the birth of
a neutron star \cite{HerreraMartinez98a,HerreraMartinez98b}. It leads, under
some circumstances \cite{Martinez96}, to densities and pressures given rise
to an equation of state similar to the Bethe-B\"{o}rner-Sato newtonian
equation for nuclear matter \cite{Demianski85,ShapiroTeukolsky83}.

The energy density and radial pressure for this static space-time are
assumed to be 
\begin{eqnarray}
\rho \left( r\right) &=&\rho _c\left[ 1-k\frac{r^2}{a^2}\right] \text{ , }
\,\,\left( 0\leq k\leq 1\right) \,,\qquad {\rm and}\qquad \\
P\left( r\right) &=&P_c\left( 1-\frac{2m\left( r\right) }r\right) \left( 1- 
\frac{r^2}{a^2}\right) ^n\text{ , }\,\,\qquad {\rm with}\qquad n\geq 1\,;
\end{eqnarray}
where $\rho _c$, $P_c$ and $a$ are the energy density, radial pressure and
the surface of the sphere, respectively. Finally, $k$ is an arbitrary
constant which may take values between zero and one and a constant parameter 
$n\geq 1$.

\subsection{The modelling performed}

In applying the HJR method to the Gokhroo and Mehra model \cite
{GokhrooMehra94}, we assume that the effective variable $\tilde{\rho}$ has
the same $r$ dependence as $\rho \left( r\right) $, i.e. 
\begin{equation}
\tilde{\rho}\left( u,r\right) =\tilde{\rho}_c\left( u\right) \left[ 1-K(u)%
\frac{r^2}{a^2}\right] \,,\,\,\,\,\left( 0\leq K\left( u\right) \leq
1\right) \,.  \label{mirho}
\end{equation}
In the above equation, $K\left( u\right) $ is a function that may take
values between zero and one only \cite{Martinez96}, and the energy density
at the center of the configuration, $\tilde{\rho}_c\left( u\right) ,$
coincides with the $\rho _c$ in the static case.

It is necessary to relate the unknown functions $K\left( u\right) $ and $%
\tilde{\rho}_c\left( u\right) $ to the surface functions $A(u)$, $\Omega (u)$
, $F(u)$ and $L(u)$. Thus, we evaluate the expression (\ref{eme}) at the
surface of the configuration: 
\begin{equation}
{m}_a=M\Rightarrow \frac{4\pi A^3\tilde{\rho}_c(u)}{15}\left( 5-3K(u)\right)
=\frac A2\left( 1-{\rm C}\left( u\right) \right) \,.
\end{equation}
Therefore, from the above equation it is possible to find $\tilde{\rho}_c:$ 
\begin{equation}
\tilde{\rho}_c=\frac{15\left( 1-{\rm C}\left( u\right) \right) }{8\pi
A^2\left( 5-3K(u)\right) }\,,
\end{equation}
and the expression for $K\left( u\right) $ is obtained from (\ref{omega}),
namely 
\begin{equation}
K\left( u\right) =\frac 53\,\frac{4\Omega -3}{8\Omega -5}=\frac 53\,\frac{
1+3\,\omega _a}{3+5\,\omega _a}\,.  \label{k}
\end{equation}
Thus, the effective energy density is written as 
\begin{equation}
\tilde{\rho}(r,u)=\frac{\left[ 1-{\rm C}\left( u\right) \right] r^2}{16\pi
A^4\left( 2\Omega -1\right) }\left[ 5\left( 3-4\Omega \right) +3\frac{A^2}{
r^2}\left( 8\Omega -5\right) \right] \ ,  \label{gokdens}
\end{equation}
and from (\ref{globalefect2}) the effective pressure, can be computed as 
\begin{equation}
\tilde{P}(r,u)=\frac{\left[ 1-{\rm C}\left( u\right) \right] r^2}{16\pi
A^4\left( 2\Omega -1\right) }\left[ 3\left( 3-4\Omega \right) +\frac{A^2}{%
r^2 }\left( 8\Omega -5\right) \right] \,.  \label{gokpres}
\end{equation}
Notice that the function ${\cal B}(u)$ should vanished in order to fulfil
the matching condition (\ref{condisup}), i.e. ${\cal B}(u)=0,\, \forall \,u$.

Finally, by virtue of (\ref{eme}) and (\ref{beta}) the metric functions read 
\begin{eqnarray}
{m}\left( u,r\right) &=&\frac{r^3\left( 1-{\rm C}\left( u\right) \right) }{%
4A^4\left( 2\Omega -1\right) }\left[ r^2\left( 3-4\Omega \right) -A^2\left(
5-8\Omega \right) \right] \,,\qquad {\rm and}\qquad  \label{gokeme} \\
\beta (r,u) &=&\frac 12\ln \left[ {\rm C}\left( u\right) \right] -\frac 12
\ln \left[ 1-\frac{2{m}(r,u)}r\right] \,.  \label{gokbeta}
\end{eqnarray}
Next section is devoted to studied the physical contribution of the {\it %
NLES } to several collapsing models emerging from the above effective
variables (equations (\ref{gokdens}) and (\ref{gokpres})) and from the
corresponding metric functions (\ref{gokeme}) and (\ref{gokbeta}).

\subsection{The {\it Systems of Surface Equations}}

At this point two family of models will be considered. The first family
emerges with ${\rm C}=const.$ which, by equation (\ref{condsup}), clearly
represents a configuration with a constant gravitational potential at the
surface, i.e. 
\begin{equation}
h_a=1-2\frac MA=F={\rm C}=const.\,.  \label{hconst}
\end{equation}
The second family of models is the most general one. It is assumed that $%
{\rm C}$ is a function of the time like coordinate, i.e. ${\rm C}={\rm C}
(u). $

In order to understand the contribution of the {\it NLES} to the
hydrodynamic scenario during the collapse, we shall compare our models with 
{\it NLES} to the most similar standard Gokhroo-Mehra-Mart\'{\i}nez model 
\cite{Martinez96}.

In the first case, ${\rm C}=const,$ the {\it system of surface equations} is 
\begin{eqnarray}
\dot{A} &=&{\rm C}\left( \Omega -1\right) \,,  \label{esconst1} \\
L &=&-\frac 12{\rm C}\left( \Omega -1\right) \left( 1-{\rm C}\right)
\,,\qquad {\rm and}\qquad  \label{esconst2} \\
\dot{\Omega} &=&\frac{2\,\Omega -1}{A\left( 1-{\rm C}\right) }\left[ L-\dot{%
A }\left( 1-{\rm C}\right) \right] -\frac{\Omega -1}{2A}\left[ 1-11\,{\rm C}
-4\,{\rm C}\,\left( 8\Omega -9\right) \,\Omega \right] \,.  \label{esconst3}
\end{eqnarray}
This system of nonlinear ordinary differential equations can be solved
(numerically) for a given set of initial values for $A\left( u\right) $, $%
F\left( u\right) $ and $\Omega \left( u\right) $.

As we have stressed above, the reference model will be the
Gokhroo-Mehra-Mart\'{\i}nez model \cite{Martinez96} with a constant
gravitational potential at the surface of the configuration. In this model
the only different equation is the third surface equation. Thus, the {\it %
system of surface equations} is conformed by equations (\ref{esconst1}), (%
\ref{esconst2}) and by the equation (4.28) in reference \cite{Martinez96}
assuming $\dot{F}=0$. Notice that, in this case, the luminosity profile,$%
L\left( u\right)$, emerges as a consequence of the condition that the
gravitational potential at the surface of the configuration be constant.

In the case ${\rm C}={\rm C}(u),$ the {\it system of surface equations} is
given by 
\begin{eqnarray}
\dot{A} &=&{\rm C}(u)\left[ \Omega -1\right] \,,  \label{esvar1} \\
\dot{{\rm C}}(u) &=&\frac{2L+{\rm C}(u)\left[ 1-{\rm C}(u)\right] \left[
\Omega -1\right] }A\,,\qquad {\rm and}\qquad  \label{esvar2} \\
\dot{\Omega} &=&\frac{2\,\Omega -1}{A\left( 1-{\rm C}(u)\right) }\left[ L- 
\dot{A}\left( 1-{\rm C}(u)\right) -\frac{A\,\dot{{\rm C}}(u)}{2\,{\rm C}(u)}
\right] -\frac{\Omega -1}{2A}\left[ 1-11\,{\rm C}(u)-4\,{\rm C}(u)\,\left(
8\Omega -9\right) \,\Omega \right] \,.  \label{esvar3}
\end{eqnarray}

The set of initial condition and parameters for three models are

\begin{center}
\begin{tabular}{|rl||rl|}
\hline
$M(0)=$ & {$1.0M_{\odot }$} & {$\rho _c(0)=$} & $2.2\times 10^{14}\,\,\,g\
\,cm^{-3}$ \\ 
$A(0)=$ & {$10\ $} & {$\rho _a(0)=$} & $9.8\times 10^{13}\,\,\,g\,\ cm^{-3}$
\\ 
$\Omega (0)=$ & {$1-\epsilon $} & {$\;a(0)=$} & $1477\,0\,\,m$ \\ 
$K(0)=$ & {$5/9$} & {$z_a(0)=$} & $0.118$ \\ 
${\rm C}(0)=$ & {$4/5$} & {$\omega _a\left( 0\right) =$} & $-1.0\times
10^{-10}\,c$ \\ \hline\hline
\end{tabular}
\end{center}

Concerning the above set, it should be pointed out that:

\begin{itemize}
\item  All of them correspond to typical values for young neutron stars.

\item  Because of (\ref{adimen2}), $\Omega =1$ corresponds to {$\omega _a=0$
. From the above {\it systems of surface equations,} it is straightforward
to verify that this initial velocity lead to static models.}

\item  {In order to obtain collapsing configurations the initial surface
velocity has been perturbed with an $\epsilon \approx $}$10^{-10}${.}

\item  In our simulations, we have imposed that the energy conditions for
imperfect fluid be satisfied. In addition, the restrictions $-1<\omega <1$
and $r>2{m}\left( u,r\right) $ at any shell within the matter configuration
are also fulfilled.

\item  For ${\rm C}={\rm C}(u)$ :

\begin{itemize}
\item  we do not need any perturbation in the velocity to start the collapse.

\item  The luminosity profile, $L\left( u\right) $, has been provided as a
Gaussian pulse centered at $u=u_p$ 
\begin{equation}
-\dot{M}=L=\frac{\Delta M_{rad}}{\lambda \sqrt{2\pi }}\exp \frac 12\left( 
\frac{u-u_p}\lambda \right) ^2,  \label{eq:mpunto}
\end{equation}
with $\lambda $ the width of the pulse and $\Delta M_{rad}$ the total mass
lost in the process. This family of models has been simulated using 
\begin{equation}
\Delta M_{rad}=10^{-6}\,M(0),\qquad \lambda =5,\qquad u_p=100\ {\rm ms.}
\label{pulset}
\end{equation}
\end{itemize}
\end{itemize}

\subsection{The Evolution of the physical variables}

The evolution of the physical variables are presented as functions of the
standard {\it Schwarzschild time. }The relationship between the usual
Schwarzschild coordinates, $(T,R,\Theta ,\Phi )$, and the previously
mentioned Bondi's radiation coordinates can be expressed as 
\begin{equation}
u=T-\int \frac r{e^{2\beta }(r-2{m}(u,r))}\;dr,\hspace{0.5cm}\theta =\Theta ,%
\hspace{0.5cm}r=R\hspace{0.5cm}{\rm and}\hspace{0.5cm}\phi =\Phi \;.
\label{eq_schcoord}
\end{equation}

Figure 2 displays the evolution of the boundary of the surface for the
Gokhroo-Mehra model with a {\it NLES }(plate (a)) and with constant
gravitational potential at the surface (plate (b)). None of them tend to any
asymptotic equilibrium configuration as was reported earlier \cite
{Martinez96}. It is apparent from these figures that our models with a {\it %
NLES} lead to a softer configurations than those without this particular
equation of state. Our matter configuration starts to collapse earlier and
evolves deeper than the reference Gokhroo-Mehra-Mart\'{\i}nez model with $%
\dot{F}=0$. In addition, because of (\ref{hconst}) $M\propto A$ , the models
with a {\it NLES} radiates more energy than those without it.

The profiles of the physical variables ($\rho $, $P_r$, $q$ , and $\omega /c$
) for three different times, are sketched in Figures 3 and 4 (plates (a)
through (d)). Figure 3 displays the profiles for the Gokhroo-Mehra model
with a {\it NLES} and Figure 4 Gokhroo-Mehra-Mart\'{\i}nez model with
constant gravitational potential at the surface. It can be appreciated from
these figures that the changes in time of the density and pressure (Figure
3, plates (a) and (b)) of the configuration with the {\it NLES} are more
significant than those observed in Figure 4, (plates (a) and (b)). In fact,
from Figure 3 (b) it is clear that the pressure rises at the inner core
while it diminishes at the outer mantle. This can be understood if we recall
that the total radial pressure $P_r$ collects the contribution of both the
hydrodynamic and radiation pressures and that it is at the outer layers
where the radiation pressure becomes more significant in time. Now,
concerning plates (b) and (c), we can see that around the zone of $r=5.5$
Kms., the total pressure remains constant and coincides with the maximum of
the energy flux density. It is clear that the energy flux density in models
with a {\it NLES} (Figure 3, plate (c)) is more intense than that observed
in the Gokhroo-Mehra-Mart\'{\i}nez reference model (Figure 4, plate (c)).
Finally notice that in Figures 4(c) and 4(d), those shells that have a
strong energy flux density also have the higher speed. This situation
contrasts with the difference that can be observed in the Figure 3, (plates
(c) and (d)) between the maximum of the energy flux density (around $r=5.5$
Kms.) and the maximum in the velocity at the $r=9.2$ Kms. This difference of
about $4$ Kms. remains constant in time during the collapse. This effect
also appears in the {\it NLES} with variable gravitational potential at the
surface (Figure 5, plates (c) through (d)), and is not present in the
corresponding Gokhroo-Mehra-Mart\'{\i}nez reported in \cite{Martinez96}. It
is worth mentioning that, because of our assumption of the diffusion regime
approximation for the radiation flux, we expect that hydrodynamics and
radiation should be closely related and this is clear in the
Gokhroo-Mehra-Mart\'{\i}nez models. Therefore, it seems that the shift
between the radiation maxima and the velocities could be thought of as the
main effect emerging from the integral contribution presented in the
nonlocal equations of state (\ref{global}) and (\ref{global2}).

Finally, in plates (e) and (f) in Figure 4 particular hydrodynamic
discontinuities, that should be mentioned, are displayed. These plates zoom
in time of the corresponding plates (c) and (d) of the same figure. As it
can be appreciated from these plates, as the energy flux rises there is a
wave front propagating outward. The discontinuity is present not in the
physical variable (either the energy flux or the matter velocity) but in its
derivative. This wave front, driven by the radiation pressure, can be
associated to a hypersurface traveling with the velocity of sound \cite
{Taub83}. It is clear that the shell where the discontinuity of the
derivative of the energy flux takes place coincides with the corresponding
surface where the discontinuity in the derivative of the velocity is
identified and this coincidence is consistent with the radiation regime that
we have assumed.

\section{Concluding Remarks}

We have shown that fluids satisfying nonlocal equations of state (either (%
\ref{global}) or (\ref{global2})) could represent feasible bounded matter
configurations and some of them could have significant geometric properties.

Some answers to the questions that motivate the present work have been
explored:

\begin{itemize}
\item  Assuming (\ref{alfa}) and ${\rm C}$ as a constant parameter in (\ref
{condicion}), we have found that spherically symmetric space-times described
by (\ref{bondisch3}), which represent fluids satisfying a {\it NLES,} have a%
{\it \ CKV} of the form (\ref{vectorexpl}).

\item  The anisotropic imperfect fluid, emerging from the metric (\ref
{bondisch3}), fulfills the corresponding energy conditions.

\item  Bounded gravitational sources can materialize from this type of
fluids and meaningful hydrodynamic scenarios can be obtained.

\item  Although the simulations for collapsing configuration where carried
out for a very short period of time, it seems that the radiating spheres we
have considered could represent some initial phases in the evolution of
compact objects.
\end{itemize}

More investigations on this curious equation of state are needed to extend
our understanding of the possible collective effects that are clear in
either (\ref{global}) or (\ref{global2}). But, as the main conclusion of the
present work it can be asserted that {\it Nonlocal Equations of State} could
represent satisfactory candidates to describe the behavior of matter at
supranuclear densities in General Relativity.

\section{Acknowledgments}

We are very grateful to L. Herrera Cometta for his helpful advices and
criticisms. We are also indebted to J. Fl\'{o}rez L\'{o}pez for pointing out
us the relevance of nonlocal theories in modern classical continuum
mechanics. Two of us (HH and LAN) gratefully acknowledge the financial
support of the Consejo de Desarrollo Cient\'{\i}fico Human\'{\i}stico y
Tecnol\'{o}gico de la Universidad de Los Andes, under project C-720-95-B and
through the Programa de Formaci\'{o}n e Intercambio Cient\'{\i}fico (Plan
II).

\begin{center}
{\bf Figure Captions}
\end{center}

\begin{description}
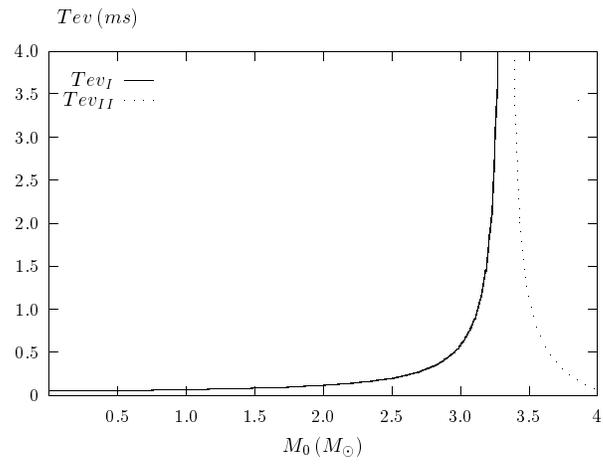
\item[Figure 1]  The total evolution times $Tev_I$ and $Tev_{II}$ are
sketched as function of the initial mass. In both cases we have assumed 10
Kms. as the value for the initial radius of the collapsing configurations.

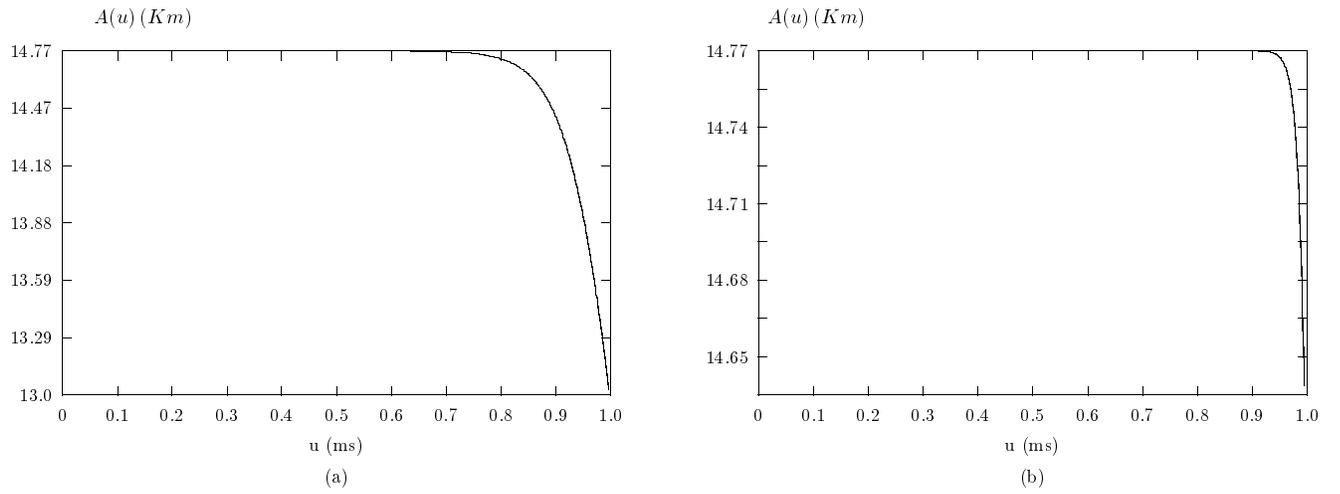
\item[Figure 2]  Evolution of the boundary radius of the matter
configuration satisfying {\it NLES}, (plate (a)) and for the standard
Gokhroo and Mehra model with constant gravitational potential at the surface
constant gravitational potential at the surface (plate (b)).

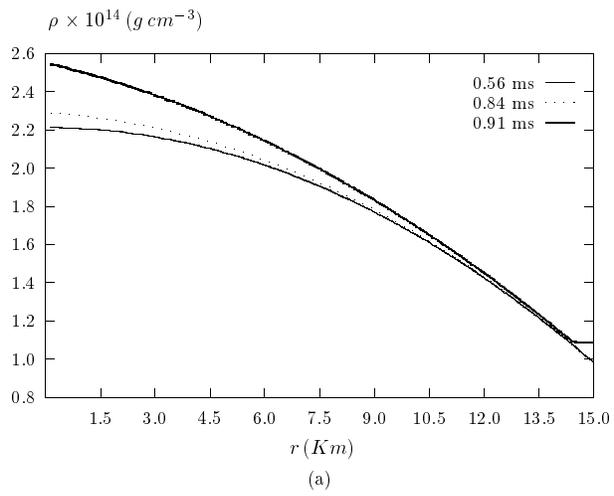
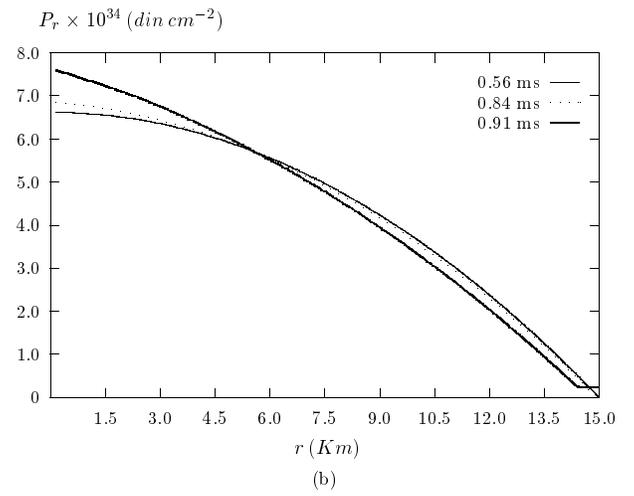
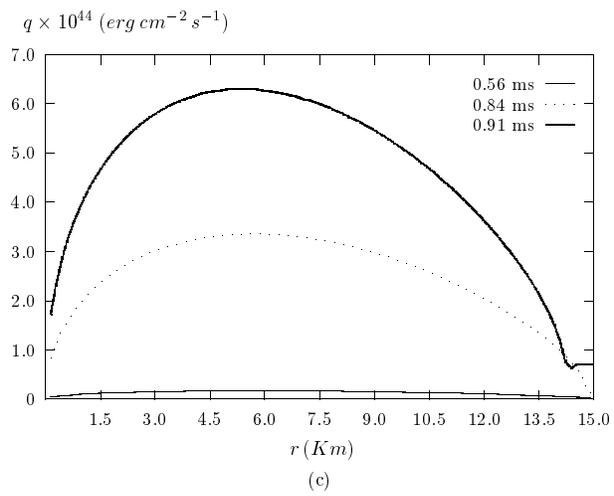
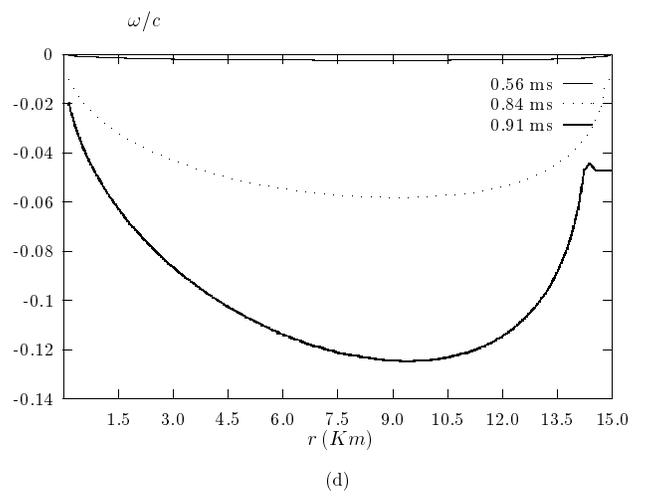
\item[Figure 3]  Evolution of matter variables for the Gokhroo and Mehra
model with a {\it NLES}. Plates (a) through (d) represent the profiles for
the density, pressure, energy flux density and mass velocity, respectively
at three different times.

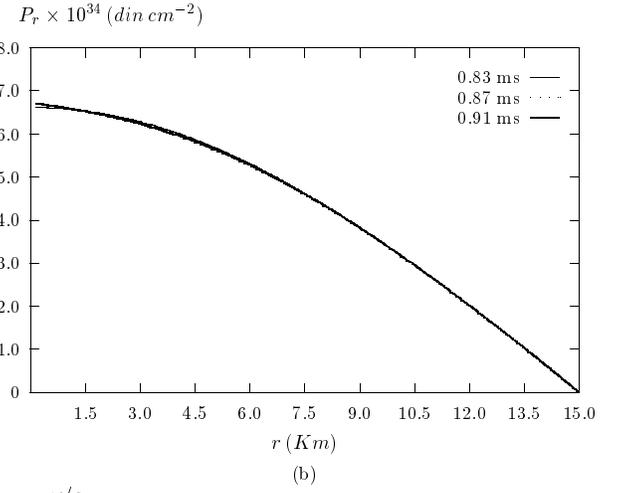
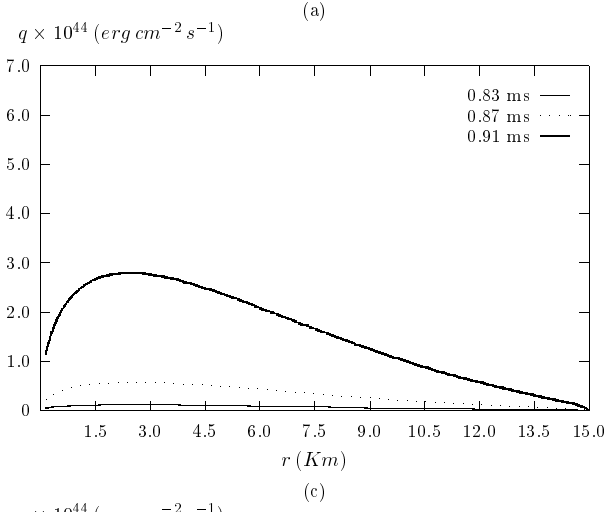
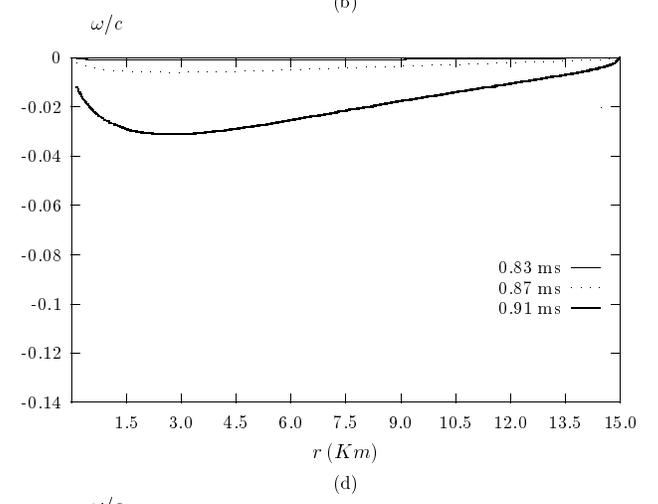
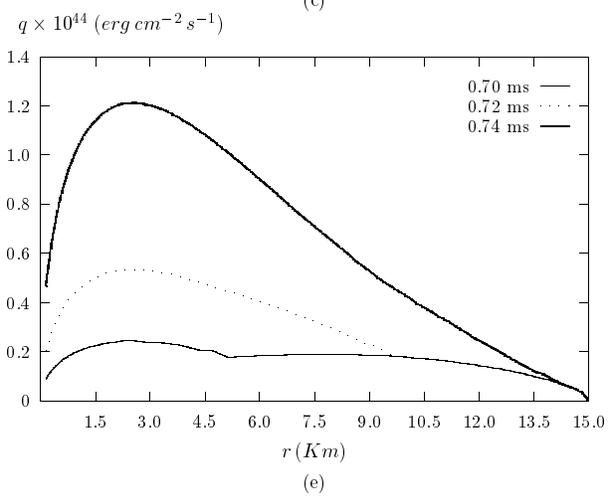
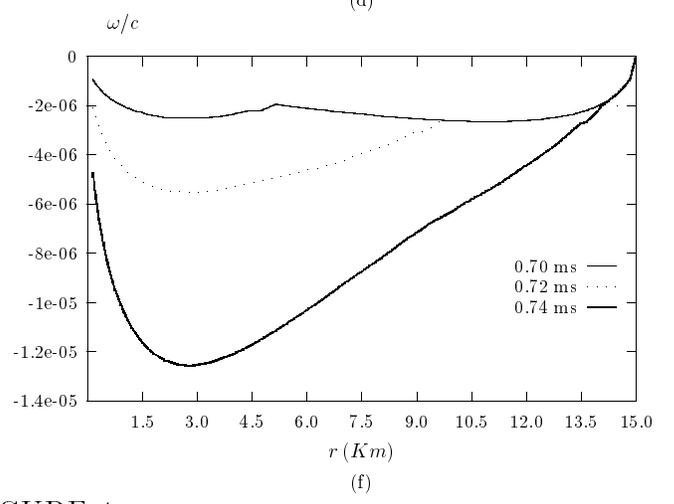
\item[Figure 4]  Evolution of matter variables for the standard Gokhroo and
Mehra model with constant gravitational potential at the surface. Plates (a)
through (d) represent the profiles for the density, pressure, energy flux
density and mass velocity, respectively at three different times. Plates (e)
and (f) are a zoom in time to the corresponding plates (c) and (d),
respectively.

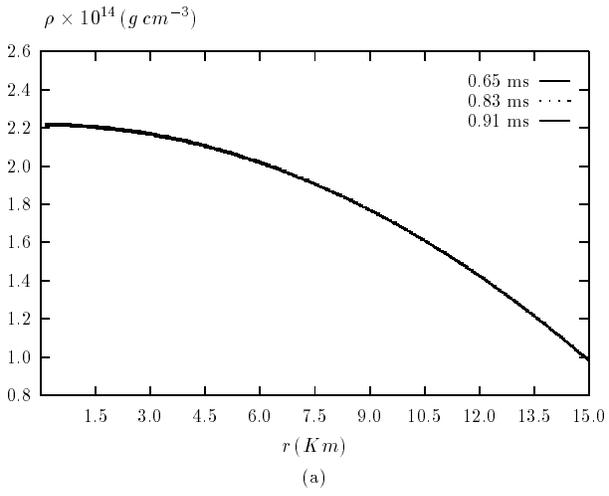
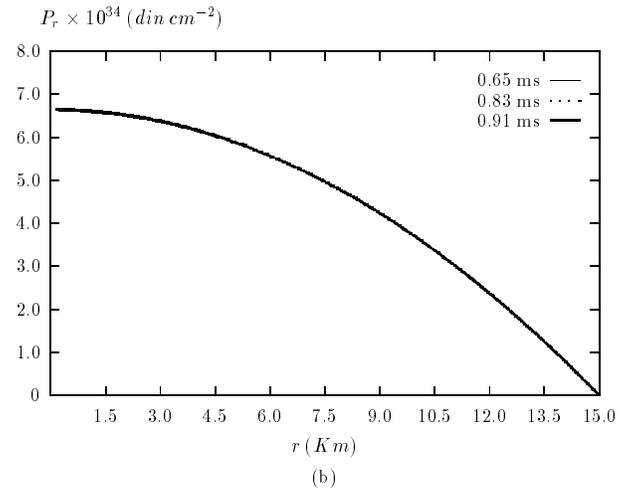
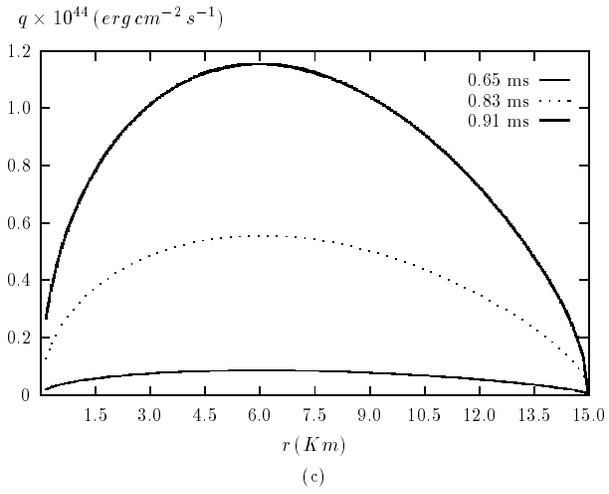
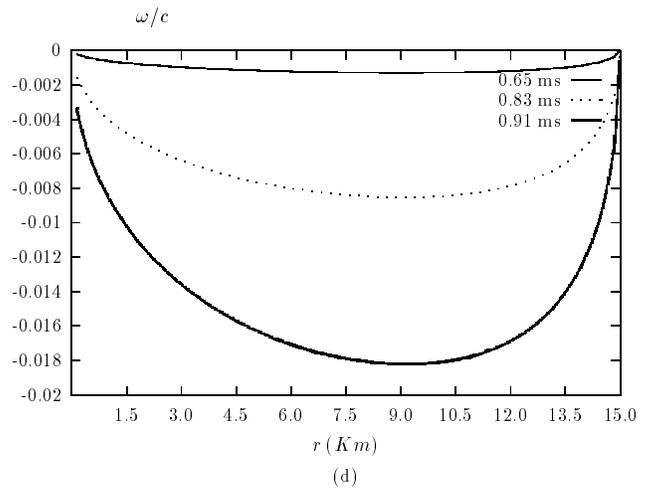
\item[Figure 5]  Evolution of matter variables satisfying {\it NLES }with $%
F(u)=1-\frac{2M(u)}{A(u)}={\rm C}(u).$ Plates (a) through (d) represent the
profiles for the density, pressure, energy flux density and mass velocity,
respectively at three different times.
\end{description}

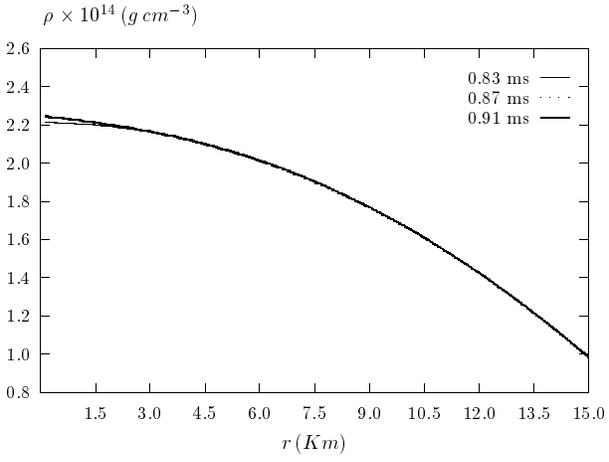
\end{document}